# Large Contribution of Quasi-Acoustic Shear Phonon Modes to Thermal Conductivity in Novel Monolayer Ga$_2$O$_3$

Gang Liu[1#*], Zhaofu Zhang[2#], Hui Wang[1], Guo-Ling Li[3], Jian-Sheng Wang[4] and Zhibin Gao[5*]

[1]School of Physics and Engineering, Henan University of Science and Technology, Luoyang 471023, People's Republic of China

[2]Department of Engineering, Cambridge University, Cambridge CB2 1PZ, United Kingdom

[3]Chemistry and Chemical Engineering Guangdong Laboratory, Shantou 515063, People's Republic of China

[4]Department of Physics, National University of Singapore, Singapore 117551, Republic of Singapore

[5]State Key Laboratory for Mechanical Behavior of Materials, Xi'an Jiaotong University, Xi'an 710049, China.

## Abstract

Bulk gallium oxide (Ga$_2$O$_3$) has been widely used in lasers, dielectric coatings for solar cells, deep-ultraviolet transistor applications due to the large band gap over 4.5 eV. With the miniaturization of electronic devices, atomically thin Ga$_2$O$_3$ monolayer has been unveiled recently, which features an asymmetric configuration with a quintuple-layer atomic structure. The superior stability, the strain-tunable electronic properties, high carrier mobility and optical absorption indicate the promising applications in the electronic and photoelectronic devices. However, the strict investigation of lattice thermal conductivity ($\kappa_L$) of 2D Ga$_2$O$_3$ is still lacking, which has impeded the widespread use in practical applications. Here, we report the computational discovery of low $\kappa_L$ with a value of 10.28 W m$^{-1}$ K$^{-1}$ at 300 K in atomically thin Ga$_2$O$_3$. Unexpectedly, two quasi-acoustic shear phonon modes contribute as high as 27% to the

[#]These authors contributed equally to this work.

*Corresponding author: Gang Liu, Email: liugang8105@gmail.com

*Corresponding author: Zhibin Gao, Email: zhibin.gao@xjtu.edu.cn

$\kappa_L$ at 300 K, leading to 37% contribution of optical phonon modes, much larger than many other 2D materials. We also find that the quasi-acoustic shear mode can emerge in the system without van der Waals interactions. This work provides new insight into the nature of thermal transport in non-van der Waals monolayer materials and predicts a new low $\kappa_L$ material of potential interest for thermal insulation in transistor applications.

## Introduction

Wide band gap semiconductor $Ga_2O_3$ has been attracting significant attention in recent years for the optoelectronic and power electronic applications [1]. With an ultrawide band gap of about 4.8 eV [2], it has a high breakdown electric field of about 9 MV cm$^{-1}$, making it attractive for high voltage device applications [3,4]. It also possesses unique ultraviolet (UV) transparency, implying potential application for novel UV optoelectronics [5,6]. Besides the electrical and optical properties, anisotropic thermal conductivity properties of β-$Ga_2O_3$ are also investigated both experimentally and theoretically [7,8].

With the scaling-down of electronic devices, it is becoming necessary to study the low-dimensional phases of traditional bulk semiconductors. Since graphene was successfully synthesized in 2004 [9], two-dimensional (2D) materials have been the focus of scientific researches, such as transition metal dichalcogenides (TMDs), group-III, -V, and -VI monolayers [10-20]. Compared with the bulk counterparts, 2D materials attract much research attention and exhibit interesting and outstanding properties such as strain-tunable band gap and high surface-area-to-volume ratio, benefiting to their wider application ranges [21,22]. Though the physical and chemical properties of bulk $Ga_2O_3$ have been studied in-depth [1-8], the 2D $Ga_2O_3$ still needs to be explored further.

Very recently, the novel 2D $Ga_2O_3$ monolayer with 2D α-$In_2Se_3$ geometry is proposed and investigated by us using first-principles calculations [23]. With an excellent dynamic and thermodynamic stability, the $Ga_2O_3$ monolayer is found to be a semiconductor with a wide indirect band gap of 3.16 eV. It has a high electron mobility of about 5000 $cm^2V^{-1}s^{-1}$, which can further increase to 7000 $cm^2V^{-1}s^{-1}$ by hybridization. The asymmetric configuration spontaneously introduces an intrinsic dipole within the quintuple-layer, boosting the separation of photon-excited carriers. Moreover, outstanding optical absorption ability is identified, which can be effectively tuned by strain engineering [23]. These outstanding properties suggest the novel $Ga_2O_3$ monolayer has great application potentials for low-dimensional optoelectronic and power electronic device. Furthermore, the intrinsic built-in field feature contributes to further applications in energy conversion such as photocatalytic water splitting or gas sensors. However, the thermal transport properties of this novel $Ga_2O_3$ monolayer are not thoroughly understood yet.

In a practical electronic device, much power is dissipated in power switching operations, causing an increase of temperature by tens or even hundreds of degrees above the ambient environment [24]. High temperatures are prone to the degradation of device performance, even destroy the device. Thus, the researches on thermal conductivity and thermal transport properties of materials are urgently required for practical applications. It should be noted that, materials with low thermal conductivity can be used as a heat insulator in practical applications, while the ones with high thermal conductivity can be used as heat dissipation materials. Thus, a thorough understanding of thermal transport properties for $Ga_2O_3$ monolayer is of technological importance.

In this work, the thermal conductivity $\kappa_L$ and thermal transport properties of $Ga_2O_3$ monolayer are systemically investigated by first-principles calculations based on the Boltzmann transport equation (BTE). It is found the monolayer has an in-plane isotropic $\kappa_L$ of 10.28 W $m^{-1}$ $K^{-1}$ at room temperature, lower than the bulk β-$Ga_2O_3$ (16 ~ 21 W $m^{-1}$ $K^{-1}$) [7, 8]. The contributions of phonon branches to total $\kappa_L$ are investigated, showing a surprisingly large proportion of 38% for all-optical branches at 300 K, which is a quite large proportion among 2D materials. Furthermore, we carefully examine the harmonic and anharmonic properties of Ga$_2$O$_3$ monolayer, to unveil the underlying physical mechanisms of the significant contribution of optical modes. The significant contribution of optical modes can be attributed to the quasi-acoustic branches with low frequency, which disperse similarly to acoustic ones. The emergence of these low-frequency quasi-acoustic shear modes results from the relatively weaker interactions within the quintuple atom layers. We investigate the effect of **quasi-acoustic shear mode** on thermal transport in materials **without** van der Waals interaction. Finally, the boundary and size effects are also studied.

## Computational and Theoretical Methods

The first-principles calculations are performed using the Vienna ab initio simulation package (VASP) [25, 26], based on density functional theory (DFT). The exchange-correlation functional is chosen in the form of the Perdew–Burke–Ernzerhof (PBE) [27]. A plane-wave basis set is employed with a kinetic energy cutoff of 600 eV, 50% higher than the maximum recommended cutoff for the pseudopotentials. The energy convergence value in structure optimization is selected as $10^{-8}$ eV and the maximum Hellmann–Feynman force is less than $10^{-4}$ eV Å$^{-1}$, while the Monkhorst–Pack [28] k-mesh of 13 × 13 × 1 is used to sample the Brillouin zone (BZ). The vacuum space of at least 20 Å is kept along the $z$-direction, which is thick enough to avoid the interactions between periodical images.

Based on the Boltzmann transport equation (BTE), the in-plane $\kappa_L$ can be expressed by [29, 30]:

$$\kappa_{\alpha\beta} = \frac{1}{V}\sum_{\lambda} C_\lambda v_{\lambda\alpha} v_{\lambda\beta} \tau_\lambda, \qquad (1)$$

where $V$ is the volume of the cell, $\lambda$ denotes a phonon mode with different wave vectors **q** and branch indexes $p$, $C_\lambda$ is the heat capacity, $v_{\lambda\alpha}$ is the group velocity along the $\alpha$ direction and $\tau_\lambda$ is the relaxation time, respectively. The group velocity is expressed as:

$$v_{\lambda\alpha} = \frac{d\omega_\lambda}{dq_\alpha}. \tag{2}$$

Eq. (1) implies the $\kappa_L$ is determined by the harmonic and anharmonic properties together. Based on Eq. (1), it can be found $C_\lambda v_{\lambda\alpha} v_{\lambda\beta} \tau_\lambda$ is the contribution to $\kappa_L$ of each phonon mode $\lambda$, while the total $\kappa_L$ is the sum. Coordinating with the frequency of each phonon mode, we can obtain the relation of the contribution to the phonon frequency. Similarly, with the contribution and the branch index of each phonon mode, we can obtain the relation of the contribution to each phonon branch.

To obtain the full solution to the BTE for phonon, an iteration approach is adopted widely with the following expression [31]:

$$\tau_\lambda = \tau_\lambda^0 (1 + \Delta_\lambda), \tag{3}$$

where

$$\Delta_\lambda = \frac{1}{N} \sum_{\lambda'\lambda''}^{+} \Gamma_{\lambda\lambda'\lambda''}^{+} (\xi_{\lambda\lambda''}\tau_{\lambda''} - \xi_{\lambda\lambda'}\tau_{\lambda'}) \\ + \frac{1}{N} \sum_{\lambda'\lambda''}^{-} \frac{1}{2} \Gamma_{\lambda\lambda'\lambda''}^{-} (\xi_{\lambda\lambda''}\tau_{\lambda''} + \xi_{\lambda\lambda'}\tau_{\lambda'}) + \frac{1}{N} \sum_{\lambda'} \Gamma_{\lambda\lambda'} \xi_{\lambda\lambda'} \tau_{\lambda'}, \tag{4}$$

$$\tau_\lambda^0 = \frac{1}{N} \left( \sum_{\lambda'\lambda''}^{+} \Gamma_{\lambda\lambda'\lambda''}^{+} + \sum_{\lambda'\lambda''}^{-} \frac{1}{2} \Gamma_{\lambda\lambda'\lambda''}^{-} + \sum_{\lambda'} \Gamma_{\lambda\lambda'} \right). \tag{5}$$

Here, $N$ is the number of **q** sampling in the Brillouin zone, $\xi_{\lambda\lambda'} = \omega_{\lambda'} v_{\lambda'}^z / \omega_\lambda v_\lambda^z$. Note in the summation $\sum^\pm$, $\lambda'' = (p'', \mathbf{q} \pm \mathbf{q}' + \mathbf{K})$, while **K** is a reciprocal lattice vector. $\Gamma_{\lambda\lambda'}$ is the isotopic impurity scattering probability [32, 33]. And the possible three-phonon transition probabilities $\Gamma_{\lambda\lambda'\lambda''}^\pm$ for mode $\lambda$ with modes $\lambda'$ and $\lambda''$ can be expressed by:

$$\Gamma_{\lambda\lambda'\lambda''}^\pm = \frac{\hbar\pi}{4} \begin{Bmatrix} f_{\lambda'} - f_{\lambda''} \\ f_{\lambda'} + f_{\lambda''} + 1 \end{Bmatrix} \frac{\delta(\omega_\lambda + \omega_{\lambda'} - \omega_{\lambda''})}{\omega_\lambda \omega_{\lambda'} \omega_{\lambda''}} |V_{\lambda\lambda'\lambda''}^\pm|^2, \tag{6}$$

where $f_\lambda$ is the Bose-Einstein distribution function depending on the phonon angular frequency $\omega_\lambda$. Furthermore, $\omega_\lambda$, $\omega_{\lambda'}$ and $\omega_{\lambda''}$ should satisfy the energy conservation, while $\mathbf{q}_\lambda$, $\mathbf{q}_{\lambda'}$ and $\mathbf{q}_{\lambda''}$ satisfy the conservation of quasimomentum. The upper (lower) row in curly brackets goes with the + (−) sign for absorption (emission) processes, respectively. $V_{\lambda\lambda'\lambda''}^\pm$ is the scattering matrix element, depending on the third-order (anharmonic) interatomic force constants (IFCs) $\Phi_{ijk}^{\alpha\beta\gamma}$, expressed as:

$$V_{\lambda\lambda'\lambda''}^{\pm} = \sum_{i\in u.c.}\sum_{j,k}\sum_{\alpha\beta\gamma} \Phi_{ijk}^{\alpha\beta\gamma} \frac{e_\lambda^\alpha(i)e_{p',\pm q'}^\beta(j)e_{p'',-q''}^\gamma(k)}{\sqrt{M_i M_j M_k}}. \qquad (7)$$

Here, $M_i$ is the mass of atom $i$, and it runs over a unit cell only in the sum. However, $j$ and $k$ run over the whole system. $e_\lambda^\alpha(i)$ means the α component of the eigenvectors of phonon mode $\lambda$ for the $i$th atom.

Eq. (3) is solved numerically for $\tau_\lambda$ with an iterative approach. Neglecting $\Delta_\lambda$ in Eq. (3), we can obtain the zeroth-order solution $\tau_\lambda = \tau_\lambda^0$, which is equivalent to the relaxation time approximation (RTA) [31, 34]. Note RTA typically does not incorporate the distinction between momentum-conserving Normal processes and resistive Umklapp processes. Thus, Normal processes are also considered as resistive in RTA, always leading to lower values of $\kappa_L$. However, the full solution to the BTE can be performed within an iteration approach, leading to higher, more reasonable, and accurate results than RTA [30]. Therefore, in this work, the discussions are always based on the results of the iteration approach, unless noted especially.

In the work, the harmonic IFCs are obtained by Phonopy [35], with a supercell of 5 × 5 × 1. The meta-GGA functional SCAN is adopted for the harmonic IFCs [36, 37]. For anharmonic IFCs, the same size supercell is adopted while the interactions are taken into consideration up to the 8$^{th}$ nearest neighbours. All DFT calculations for supercells are Γ-point only as there are 125 atoms in the supercell. Then the anharmonic IFCs are extracted by thirdorder.py script, and the thermal conductivity is calculated by ShengBTE [34]. After the careful test, we chose a dense k-mesh grid of 151 × 151 × 1, to ensure the convergence of thermal conductivity.

## Results and Discussions

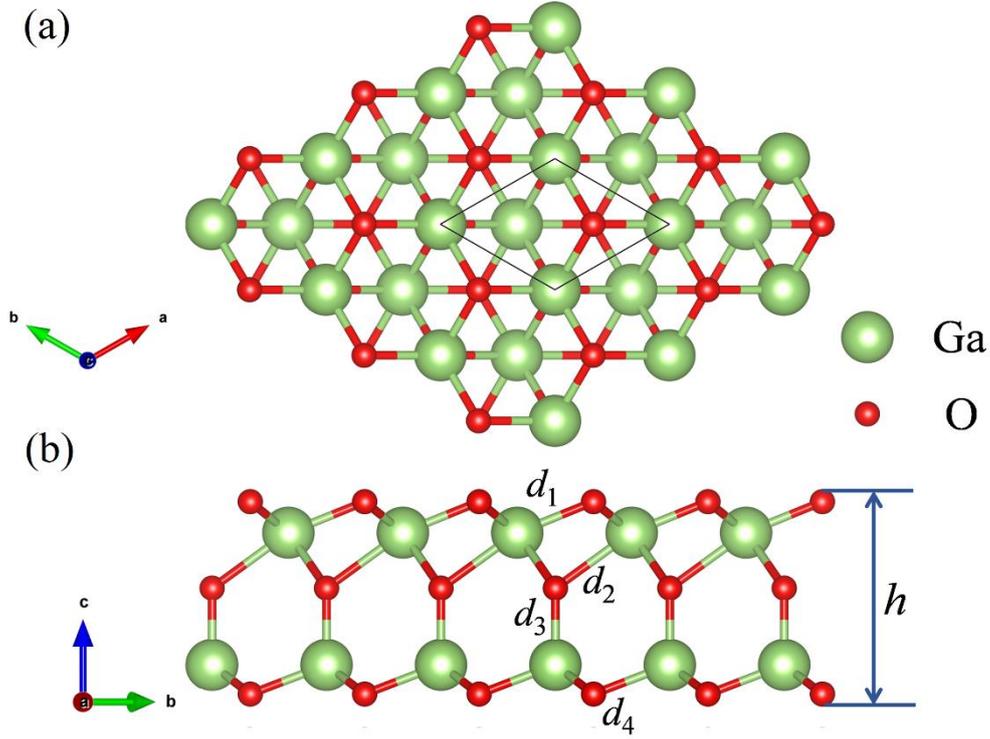

Fig. 1. Top view (a) and side view (b) of the optimized structure of the novel $Ga_2O_3$ monolayer. The primitive cell is marked by black solid lines in the top view. Note in (b) four types of Ga-O bonds are labeled as $d_1$, $d_2$, $d_3$, and $d_4$, respectively.

As shown in Fig. 1, the stable structure of the $Ga_2O_3$ monolayer belongs to *P*3*M*1 (156) symmetry group, also features an isotropic pattern in the 2D plane. The side view in Fig. 1(b) shows the stacked atomic layer in the sequence of O-Ga-O-Ga-O, forming the quintuple layer consisted of covalently bonded gallium and oxygen triangular lattices. The different Ga-O bond lengths are labeled as $d_1$, $d_2$, $d_3$, and $d_4$, which are 1.92, 2.21, 1.80, and 1.91 Å, respectively. The optimized lattice parameters are $a = b = 3.08$ Å, slightly larger than the previous work (3.04 Å) [23]. This is owing to the cell is relaxed using PBE functional in this work, which gives a slightly larger lattice constant than hybrid functional [38].

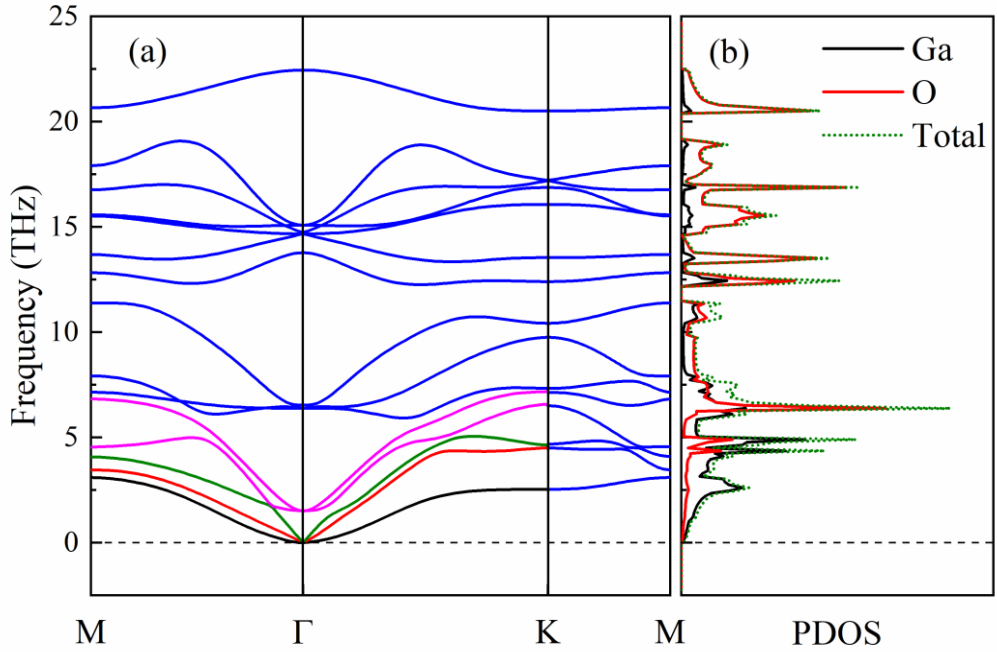

Fig. 2. Phonon dispersion (a) and PDOS (b) of Ga$_2$O$_3$ monolayer. In (a), the black, red, and green lines indicate ZA, TA, and LA modes, respectively. The quasi-acoustic optical branches are displayed by purple lines while other optical branches are represented by blue lines.

The phonon dispersions and phonon density of states (PDOS) are calculated and shown in Fig. 2(a) and (b). The stability of Ga$_2$O$_3$ monolayer is identified as there is no imaginary frequency. Since there are 5 atoms in the primitive cell, 15 phonon branches exist, including 3 acoustic and 12 optical branches. Note the out-of-plane acoustic (ZA, black curve) phonon mode is quadratic around the Γ point, which is the feature of 2D materials owing to the membrane effect. And there are other two acoustic branches: transverse acoustic (TA, red curve) and longitudinal acoustic (LA, green curve) branches, which show linear relationships with $q$ near the Γ point [39]. The unique frequency dependence of the three acoustic branches can be understood by the 2D continuum elasticity theory [40]. Moreover, the two lowest optical branches around Γ point (purple curves) are well separated from the other optic branches, named "quasi-acoustic" modes since they disperse very similarly to acoustic modes [41]. In fact, the quasi-acoustic branches are very common in many bilayers and layered materials [41-44], due to the weak layer-layer interactions. In Fig. 2(b), it is found that the Ga atoms dominate the low-frequency region because of their heavier mass, while the lighter O atoms contribute mainly in the high frequency. In the low-frequency range lower than 5 THz, there are three significant PDOS peaks round about 2.6, 4.4, and 4.9 THz. Note the quasi-acoustic modes are also within this frequency range and should contribute much to the PDOS. Larger PDOS indicates that more phonon modes can carry heat, hence more contribution to $\kappa_L$. And these low-frequency peaks are mainly related to heavier Ga atoms.

The Debye temperature $\theta_D$ is an important physical quantity related to the thermal properties. Usually, high $\theta_D$ means high thermal conductivity [45]. However, it should be noted $\theta_D$ can not determine $\kappa_L$ separately. $\kappa_L$ is affected by several harmonic and anharmonic phonon properties, and $\theta_D$ mainly reflects the magnitude of phonon group velocity $v_g$, a harmonic property. It can be obtained by $\theta_D = \hbar\omega_{max}/k_B$, where $\omega_{max}$ is the maximum of acoustic phonon frequency [46, 47]. The calculated value is 241 K with this expression, higher than 2D SnSe (87 K), β-tellurene (106 K), stanene (198 K) and 2D $SnS_2$ (233 K), but lower than 2D $MoS_2$ (278 K) and graphene (2359 K)[48-51].

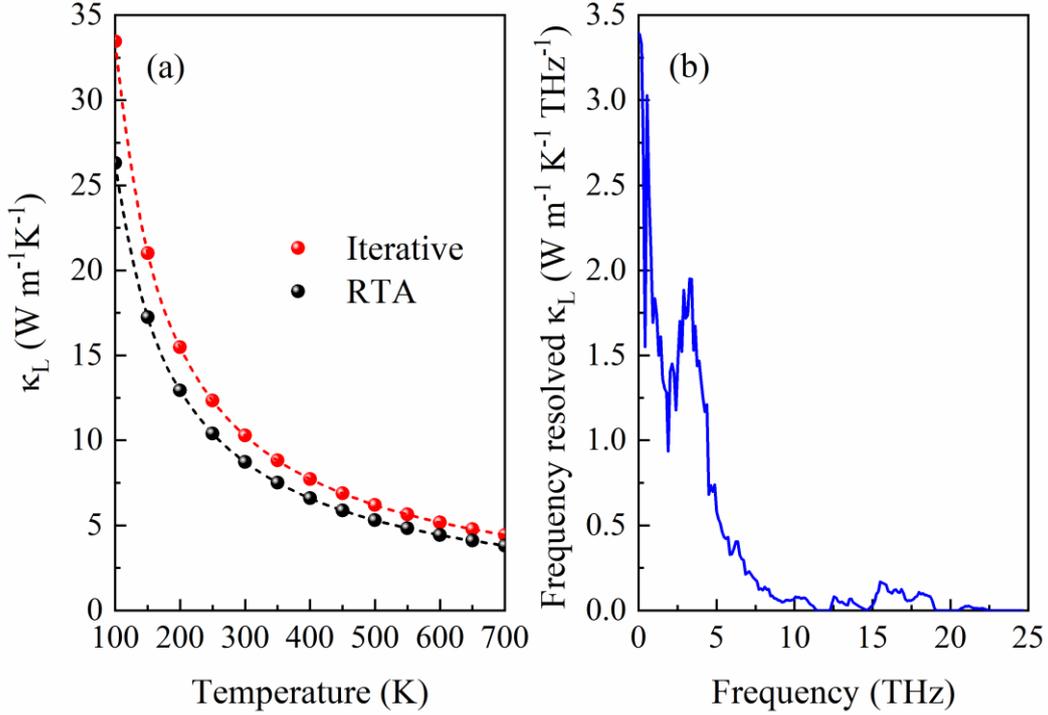

Fig. 3. (a) Calculated $\kappa_L$ of Ga$_2$O$_3$ monolayer and (b) the frequency-resolved $\kappa_L$ at 300 K. The dashed line in (a) indicates the 1/T fitting of temperature-dependent $\kappa_L$.

The calculated $\kappa_L$ of iterative method is plotted in Fig. 3(a). Note the $\kappa_L$ of Ga$_2$O$_3$ monolayer is in-plane isotropic, resulting from its in-plane isotropy of structure. It should be noted that an effective thickness should be defined to calculate the $\kappa_L$ for 2D materials. The effective thickness is 7.57 Å, with the definition of the summation of the buckling height $h$ and twice of the van der Waals radii of the outermost O atoms [47, 48, 52]. The $\kappa_L$ of Ga$_2$O$_3$ monolayer is 10.28 W m$^{-1}$ K$^{-1}$ at room temperature, lower than bulk β-Ga$_2$O$_3$ [7, 8]. To compared $\kappa_L$ with other 2D materials, we also use the thermal sheet conductance ("2D thermal conductivity") with the unit W K$^{-1}$ as it is more meaningful and physical for 2D materials [52]. Then we get the value of 7.77 nW K$^{-1}$ for Ga$_2$O$_3$ monolayer at 300 K. It is also a quite small value in 2D materials, smaller than SnS$_2$, MoS$_2$, MoSSe, and MoSe$_2$ monolayers [50, 53]. Furthermore, we found that $\kappa_L$ of Ga$_2$O$_3$ monolayer matches well with $T^{-1}$ behavior, indicating the Umklapp process of phonon scattering dominates the thermal transport [54, 55]. The RTA results are also shown in Fig. 3(a), which are lower than the results of the iterative method. For instance, the $\kappa_L$ of RTA is 8.72 W m$^{-1}$ K$^{-1}$ at 300 K. The difference between two methods is small, in the agreement of the common criteria that Normal processes usually are relevant only for materials with high $\kappa_L$ [34, 56]. It also implies the effect of Normal processes can be neglected in a rough approximation such as RTA.

To examine the contributions of phonons with different frequencies to the total $\kappa_L$, the frequency-resolved $\kappa_L$ of Ga$_2$O$_3$ monolayer at 300 K is calculated and shown in Fig. 3(b). There are two significant peaks in Fig. 3(b) locating around 0 and 3.3 THz, indicating nearby phonons contribute greatly to $\kappa_L$. It can be found most of the contributions come from phonons lower than 5 THz. As most quasi-acoustic phonons have a frequency lower than the value, it implies the optical modes should have a significant contribution to the total $\kappa_L$.

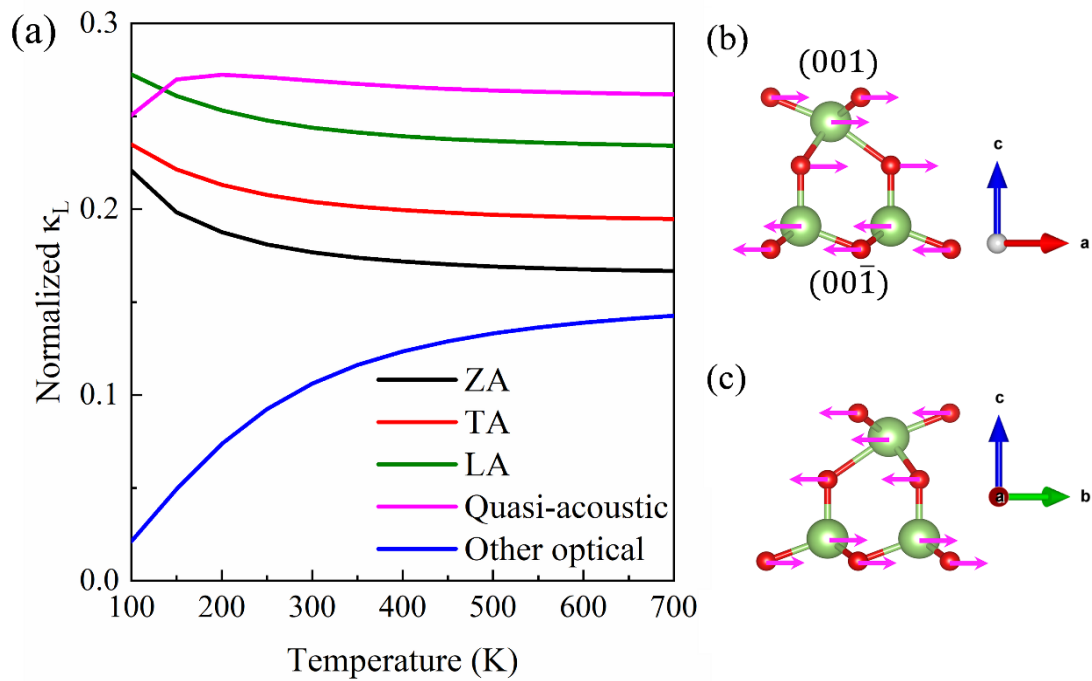

Fig. 4. (a) Normalized $\kappa_L$ of phonon modes with increasing temperature. (b) shows vibrating patterns of the lowest quasi-acoustic mode near Γ point along Γ-M direction, while the one of second-lowest quasi-acoustic mode is displayed in (c).

The normalized contribution of each phonon branch to the total $\kappa_L$ versus temperature is shown in Fig. 4(a). Note the normalizing factor is the total $\kappa_L$. The total contribution of optical branches except for the two quasi-acoustic branches, rises with increasing temperature, while the ones of three acoustic modes decline. It results from the fact that only acoustic phonons with low frequency can be activated at low temperatures, while most optical phonons with high frequency can also be activated at high temperatures. However, the contribution of quasi-acoustic modes also declines at high temperatures, as they possess low frequency similar to acoustic modes. At 300 K, the normalized $\kappa_L$ is 0.18, 0.20, and 0.24 for ZA, TA, and LA modes, whereas the sum of the two quasi-acoustic modes is 0.27. The contribution from all the quasi-acoustic modes is quite large, resulting in a percentage up to 38% from the optical branches. The vibrating patterns of the two quasi-acoustic modes are exhibited in Fig. 4(b) and (c). It should be noted that the vibrations of the quasi-acoustic branches show remarkable layered motion. Specifically, the top three-atom layers of (001) surface vibrate together, while the bottom two-atom layers of $(00\bar{1})$ surface show the out-of-phase motion compared with the top three atom layers. The relative motions of the top O-Ga-O layer and the bottom Ga-O layer are parallel to the layer plane, similar to the shear modes in bilayer/bulk transition metal dichalcogenides, as well as other layered materials [43, 44]. It should be noted that quasi-acoustic modes also contribute to $\kappa_L$ significantly in bulk $MoS_2$ [42]. Note there is no low-frequency quasi-acoustic mode that vibrates along the z-direction compared with bulk $MoS_2$. It is reported the quasi-acoustic mode along z-direction results from the symmetry of atomic layers along the same direction [43, 44]. Therefore, the atomic layers don't show symmetry along the z-direction in $Ga_2O_3$ monolayer, leading to the lacking of quasi-acoustic mode vibrating along this direction. Tough the acoustic modes contribute to the total $\kappa_L$ more than all the optical modes (62% compared with 38%), the proportion of contribution for optical modes is still much more than many 2D materials, such as graphene (1%) [29, 51], α-tellurene (10%) [57], $MoS_2$ monolayer (1.4%) [49], and stanene (2.1%) [58].

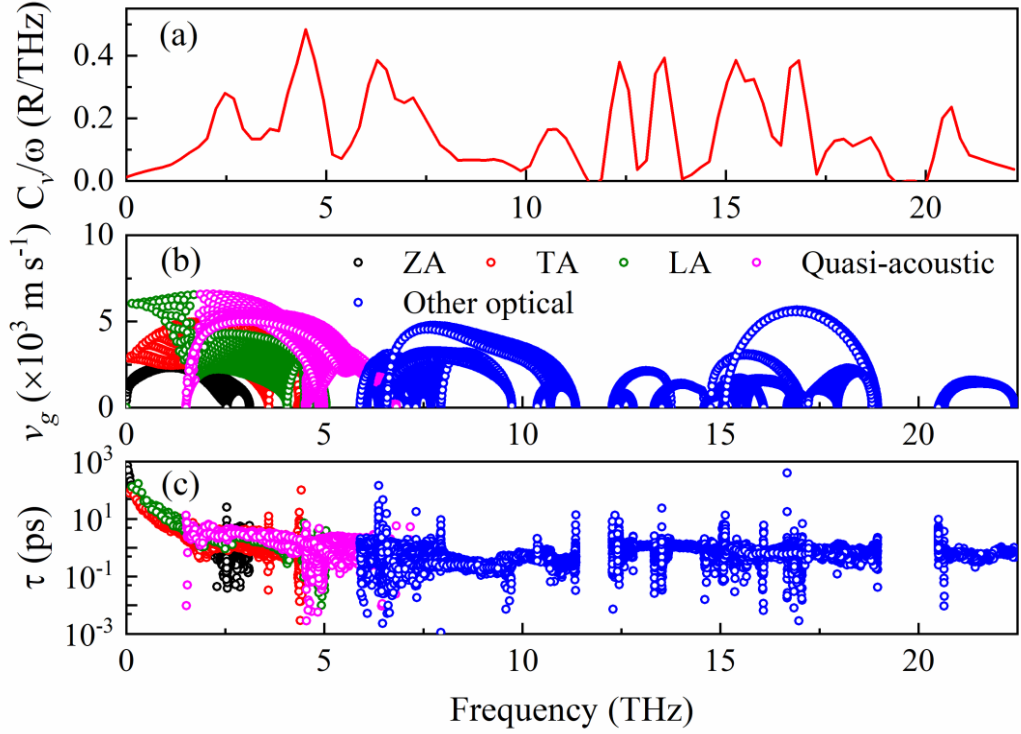

Fig. 5. Frequency resolved heat capacity (a), phonon group velocity $v_g$ (b), and relaxation time $\tau$ (c) of $Ga_2O_3$ monolayer at 300 K.

To further reveal the underlying physics for the low $\kappa_L$ and abnormal high contribution of optical modes, the heat capacity $C_v$, phonon velocity $v_g$, and phonon relaxation time $\tau$ are investigated, as shown in Fig. 5. The frequency-resolved heat capacity at 300 K is displayed in Fig. 5(a). There are several peaks of the curves in the range of low and high frequencies, indicating phonons with low frequency and high frequency have a remarkable contribution to heat capacity at room temperature. The two peaks in the low-frequency range are near 2.5 and 4.5 THz, in reasonable agreement with the results in Fig. 2(b). In fact, at 300 K where it is higher than the Debye temperature (241 K), the heat capacity $C_v$ of each phonon branch approaches the classic value $k_B$, the Boltzmann's constant. Thus, large PDOS always indicates a significant contribution to $C_v$ [59]. In Fig. 5(b), the dispersions of phonon group velocity $v_g$ are plotted. The group velocity is high in both low and high-frequency ranges. Especially, a great number of phonons with a frequency lower than 5 THz has a very high group velocity exceeding $6 \times 10^3$ m s$^{-1}$. Among the three acoustic branches, LA phonons have the highest velocity, and the value of ZA phonons is the lowest in this frequency range. It is notable that the low optical phonons, i. e., the quasi-acoustic phonons, have almost the largest group velocity in the range. It is also in good agreement with Fig. 2, where the quasi-acoustic modes disperse similarly to acoustic modes, implicating the similar phonon group velocity, based on Eq. (2). Furthermore, the optical phonons in the range of about 7~10 THz and 16~19 THz also possess high group velocity. However, these high-frequency optical phonons have little contribution to the $\kappa_L$, as shown in Fig. 3(b).

At last, the calculated phonon relaxation time $\tau$ at 300 K is displayed in Fig. 5(c). It can be found that the acoustic phonons around Γ point with very low frequency have relaxation time as long as $10^3$ ps. However, the phonon relaxation time decreases quickly with increased frequency, and it is in the order of 10 ps when the frequency is round 1 THz. Then the relaxation time declines slowly. It should be noted that most of the quasi-acoustic phonons are lower than 5 THz (Fig. 2), where the relaxation times are round the high value of 10 ps. Most phonons have a relaxation time shorter than 1 ps when the frequency is higher than 5 THz. On the whole, the relaxation time mainly determines the contributions of phonons belonging to various modes and frequency ranges to the total $\kappa_L$. Specifically, the phonons with a frequency lower than 5 THz have relaxation time at least an order higher than the ones of other phonons, combining with high velocity and heat capacity, leading to the dominating contribution to $\kappa_L$, as shown in Fig. 3(b). Among these low-frequency phonons, a large number of quasi-acoustic phonons, which also possess high relaxation time, velocity, and remarkable heat capacity, should contribute remarkably to $\kappa_L$. However, there are no optical phonons with a very low frequency lower than 1 THz, where the relaxation time is higher than $10^2$ ps. As a result, the contribution of the optical modes is still less than acoustic modes, though their contribution is pretty significant, close to 40%. Compared with other 2D materials, the $\tau$ of $Ga_2O_3$ monolayer is quite low, which is the key point leading to the lower $\kappa_L$. For instance, though the $v_g$ are close to each other, the highest value of $\tau$ is on the order of $10^3$ and $10^4$ ps for $Ga_2O_3$ and $MoS_2$ monolayer respectively, leading to much lower $\kappa_L$ in $Ga_2O_3$ monolayer [49].

It can be argued reasonably that quasi-acoustic optical modes can enhance the proportion of contribution to $\kappa_L$ for optical branches greatly, since they possess harmonic and anharmonic properties similar to acoustic ones. Furthermore, some models such as the Slack model [45], which are extensively used to estimate $\kappa_L$ of materials fast and conveniently, assume that only the acoustic phonon modes participate in the heat conduction process. It should be cautious about using these models when the quasi-acoustic branches emerge, since the quasi-acoustic branches may contribute remarkably to $\kappa_L$.

The effect of quasi-acoustic modes on $\kappa_L$ is analyzed above. It is reported that the quasi-acoustic modes usually emerge in layered materials, where the atomic layers interact through weak van der Waals interaction [41]. Now we change the bond length of $d_1$, $d_2$, $d_3$, and $d_4$ separately by displacing the atoms, to examine the corresponding potential energy surfaces. First, we move atoms with a fixed step to stretch/compress the distance of bonds that we want to study, while other bonds remain unchanged. Then the total energies of these new structures are calculated, and the corresponding potential energy surfaces can be obtained. The results are plotted in Fig. 6. The potential energy surfaces of in-layer and inter-layer interactions in bulk $MoS_2$ are also calculated and displayed in Fig. 6 for comparison. We use the second-order derivatives of these curves to measure the bond strengths, which are 57, 27, 26, and 133 eV/Å$^2$ for $d_1$, $d_2$, $d_3$, and $d_4$, respectively. Note the weakest $d_3$ bond is along the $z$-direction, while other stronger bonds are partly in-plane. Thus, the strong bonds connect the atoms in the top O-Ga-O layer of the (001) surface, forming a unit that moves together. And the same to the atoms in the bottom O-Ga layer of the (00$\bar{1}$) surface. The weakest $d_3$ bond connects the top O-Ga-O layer to the bottom O-Ga layer, resulting in the relative vibrations of the two parts (Fig. 4(b) and (c)), and the low-frequency quasi-acoustic modes emerge. Moreover, in bulk $MoS_2$ the second-order derivative for the inter-layer van der Waals interaction is 11 eV/Å$^2$, while the one for the in-layer bonds is 258 eV/Å$^2$. It can be concluded quasi-acoustic modes also appear when the in-layer interactions are much stronger than the inter-layer ones, similar to $Ga_2O_3$ monolayer. Thus, we emphasize when there exists much weaker inter-layer interaction than the one of inter-layer, low-frequency quasi-acoustic modes emerge, whereas the van der Waals interaction is not necessary for this.

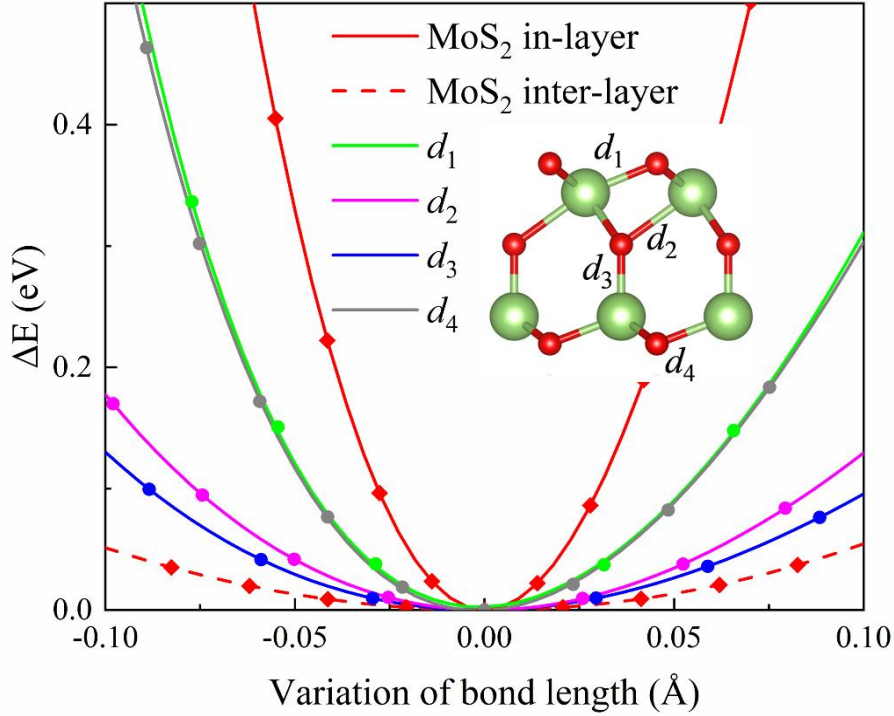

Fig. 6. Potential energy surfaces of $d_1$, $d_2$, $d_3$, and $d_4$ in 2D $Ga_2O_3$. The ones of in-layer and inter-layer interactions in bulk $MoS_2$ are also given for comparison. The inset shows the side view of $Ga_2O_3$ monolayer, where the atomic bonds $d_1$ to $d_4$ are also displayed.

In practical applications, $\kappa_L$ of materials may be significantly suppressed as all materials have finite sizes, where the additional boundary scattering can significantly affect $\kappa_L$, especially at the nanoscale. Usually, an empirical formula is used to describe the boundary scattering $\tau_\lambda^b$, which is expressed as: $\frac{1}{\tau_\lambda^b} = \frac{v_\lambda}{L}$, where $L$ means the size of a material [60]. The normalized $\kappa_L$ as a function of sample size $L$ at room temperature is shown in Fig. 7(a). The normalized $\kappa_L$ of $Ga_2O_3$ monolayer declines following an exponential function of decreasing $L$ due to the stronger boundary effect. In fact, the dependence of *L* has been experimentally verified in suspended graphene [60, 61]. The normalized $\kappa_L$ of Ga$_2$O$_3$ monolayer is 0.50 at the size of 15 nm. When the *L* is 10$^3$ nm, the normalized $\kappa_L$ is 0.95, indicating the boundary effect is very weak and can be neglected. To estimate the size effect, we also evaluate the normalized cumulative $\kappa_L$ with respect to the phonon mean free paths (MFPs) for the monolayer, as exhibited in Fig. 7(b). The phonon MFPs contribute mainly to $\kappa_L$ in the range of 1 to about 200 nm. In order to obtain the characteristic length, we introduce a single parametric function [34]:

$$\kappa_L(l \leq l_{max}) = \frac{\kappa_{max}}{1 + l_0/l_{max}}, \quad (8)$$

where $l_{max}$ and $\kappa_{max}$ are the maximum MFP and ultimately cumulative lattice thermal conductivity. Only a parameter $l_0$ needs to be determined in the expression, which is regarded as the representative MFP. It is found $l_0$ is 15 nm for Ga$_2$O$_3$ monolayer, corresponding to the positions of 50% of the total $\kappa_L$. It can be expected that the $\kappa_L$ will significantly decrease when the size is on the order of 10 nm. The size effect discussion provides useful reference and guidance for thermal management design in micro-/nano-electronic devices based on this novel Ga$_2$O$_3$ monolayer.

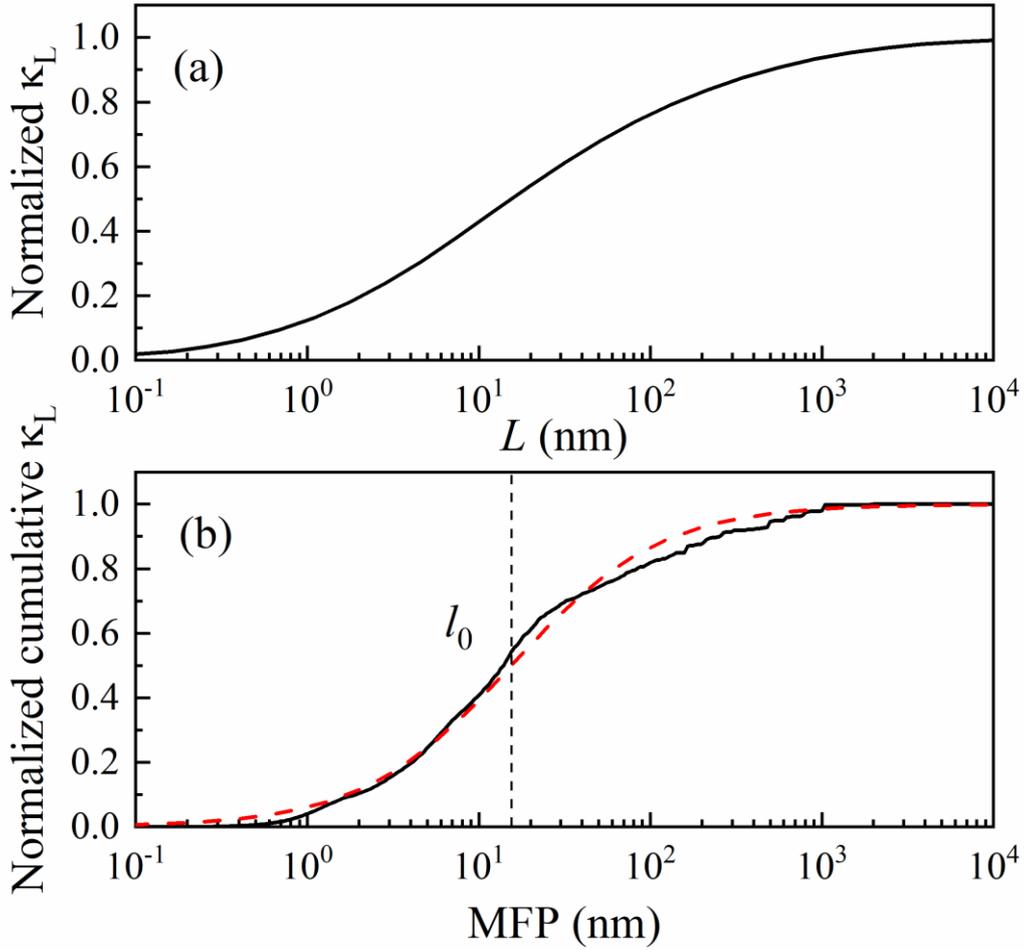

Fig. 7. Normalized $\kappa_L$ as a function of sample size $L$ (a) and the MFPs dependent normalized cumulative $\kappa_L$ at 300 K (b). In (b) the dashed red line represents the curve of the fitting, while the vertical dashed line indicates the position of $l_0$ for the Ga$_2$O$_3$ monolayer.

## Conclusion

In summary, the lattice thermal conductivity $\kappa_L$ of novel Ga$_2$O$_3$ monolayer is investigated based on first-principles calculations. Compared to its bulk counterpart (16 ~ 21 W m$^{-1}$ K$^{-1}$), the $\kappa_L$ for Ga$_2$O$_3$ monolayer is only 10.28 W m$^{-1}$ K$^{-1}$ at room temperature, which is a quite low $\kappa_L$ among various 2D materials. Though the acoustic branches dominate the thermal transport in the monolayer, optical modes contribute significantly to the total $\kappa_L$, close to 40% at 300 K. The harmonic and anharmonic phonon properties determine the $\kappa_L$ of Ga$_2$O$_3$ monolayer together. It is found the relaxation time plays a vital role in the thermal transport of Ga$_2$O$_3$ monolayer. The low-frequency quasi-acoustic phonons possess significant PDOS and heat capacity, as well as high relaxation times and phonon velocity, resulting in a considerable contribution to total $\kappa_L$. It is argued that quasi-acoustic optical modes can greatly enhance the proportion of contribution to $\kappa_L$ for optical branches. The results provide important information for the systematic understanding of the thermal transport properties of the novel $Ga_2O_3$ monolayer, as well as its future design in micro-/nano-devices for practical applications.

## Data Availability Statement

The data that support the findings of this study are available from the corresponding author upon reasonable request.

## Acknowledgments

This work was supported by the National Natural Science Foundation of China (Nos. 11974100, 1210040387). We acknowledge the strong support by HPC Platform, Xi'an Jiaotong University.